\documentclass[copyright,creativecommons]{eptcs}
\providecommand{\event}{ICE 2010} 
\usepackage{breakurl}             

\title{History-sensitive versus future-sensitive approaches to security in distributed systems}
\author{Alejandro Hernandez and Flemming Nielson
\institute{Department of Informatics, Technical University of Denmark \\ \texttt{\{aher,nielson\}@imm.dtu.dk}}}
\def\titlerunning{History- vs. future-sensitive approaches to security in distributed systems}
\def\authorrunning{A. Hernandez and F. Nielson}
\usepackage{amsmath,amssymb,enumerate,color,graphicx,subfig}

\newcommand{\TrueB}{\textbf{t\!t}}
\newcommand{\FalseB}{\textbf{f\!f}}
\newcommand{\TrueS}{{\bf true}}
\newcommand{\FalseS}{{\bf false}}
\newcommand{\textif}{\ \IF\ }

\newcommand{\Aspectbegin}{$$ \left[ \begin{array}{c}}
\newcommand{\Aspectend[1]}{\end{array} \right]#1$$}
\newcommand{\AspectEqbegin}{\left[ \begin{array}{c}}
\newcommand{\AspectEqend[1]}{\end{array} \right]#1}

\newcommand{\netpar}{\mid\mid}
\newcommand{\val}[1]{\langle\,#1\,\rangle}

\newcommand{\ppar}{\mid}

\newcommand{\OutM}[1]{{\bf out}({#1})}
\newcommand{\InM}[1]{{\bf in}({#1})}
\newcommand{\ReadM}[1]{{\bf read}({#1})}
\newcommand{\locate}[2]{{#1}\,@\,{\it #2}}

\newcommand{\nil}{{\bf 0}}

\newcommand{\YZl}{l}
\newcommand{\Fail}{{\sf fail}}

\newcommand{\Lnt}{\ell^\lambda}
\newcommand{\Ln}{\ell}

\newcommand{\Lc}{\ell}

\newcommand{\Lat}{\ell^\lambda}

\newcommand{\veck}[1]{\overrightarrow{#1}}

\newcommand{\doublebracketleft} {[\![}
\newcommand{\doublebracketright}{]\!]}
\newcommand{\db}[1]{\doublebracketleft #1 \doublebracketright}
\newcommand{\dt}[1]{[\!( #1 )\!]}

\newcommand{\grant}[1]{{\sf grant}({#1})}

\newcommand{\Inference}[2]{\begin{array}{@{}c@{}}#1\\[0em]\hline\\[-0.9em]#2\\
\end{array}}

\newcommand{\IF}{\underline{\sf if}\ }

\newcommand{\Localised}[3]{#1::^{#2}#3}
\newcommand{\LocalisedSubject}[2]{\Localised{\YZl_s}{#1}{#2}}
\newcommand{\LocalisedTarget}[2]{\Localised{\YZl_t}{#1}{#2}}
\newcommand{\w}{w}

    \newtheorem{theorem}{Theorem}[section]
    \newtheorem{lemma}[theorem]{Lemma}
      
\begin{document}
\maketitle

\begin{abstract}
We consider the use of aspect-oriented techniques as a flexible way to deal 
with security policies in distributed systems.
Recent work suggests to use aspects for analysing the future behaviour of 
programs and to make access control decisions based on this; this gives 
the flavour of dealing with information flow rather than mere access control.
We show in this paper that it is beneficial to augment this approach with 
history-based components as is the traditional approach in reference
monitor-based approaches to mandatory access control.
Our developments are performed in an aspect-oriented coordination
language aiming to describe the Bell-LaPadula policy as elegantly as
possible.
Furthermore, the resulting language has the capability of combining both history- and future-sensitive policies, providing even more flexibility and power.
\end{abstract}

\begin{section}{Introduction}
Distributed Systems are designed to manage large amounts of information, so they must be secured~\cite{Shamon} to provide confidentiality for the information managed by them. The emerging Aspect-Orientation~\cite{Aspects} field has been targeted to some security approaches~\cite{AspectK}. Recently, a framework named \textbf{AspectKB}~\cite{AspectKB} has been proposed, with which is possible to model process calculi-like distributed systems and to capture security properties in a realistic way, attaching security policies to each location and then combining the relevant security policies when an interaction between locations takes place.

The way of expressing security policies in the \textbf{AspectKB} framework refers to the traditional non-distributed information-flow~\cite{InformationFlow} style of assuring security, which statically analyses the possible behaviours of the system in order to avoid any potential misuse in the future. In \textbf{AspectKB}, this is exploited by making access control decisions dynamically, yet not considering any state of the locations but possibly some potential future behaviour.

In this paper, we shall consider a multilevel access control policy~\cite{Dieter}, the Bell-LaPadula model~\cite{BLP}, and show some complications when trying to capture such a policy in a distributed framework in general, and in particular in a framework whose security policies focus on \emph{looking to the future}, since such multilevel policies are better suited for past analysis of how the system reached its current state.

We then propose an extension to the \textbf{AspectKB} framework, allowing to express also policies that \emph{look to the past}. We do this by adding the notion of a \emph{localised state} to the locations modelled in the extended framework and allowing the security policies to access those states to make their decisions. With this, multilevel policies as the Bell-LaPadula policy can be easily captured, and we show how.

Since the original \textbf{AspectKB} framework was already intended to combine different security policies, with the extension done in this paper both policies that look to the past and policies that look to the future can be expressed and even combined. This not only benefits when trying to capture specific policies (such as the Bell-LaPadula one), but it also allows us to model every policy in its original way, providing more flexibility to the resulting extended framework.

Moreover, we shall argue that for some situations, expressing a policy in its original way could more precisely capture what is intended, and this insight would mean that our extended framework is more powerful as well. We shall start discussing this latter issue in the remainder of this Section (Subsection \ref{subsec:Limitations}). In Section \ref{sec:BLP} we present a review of the Bell-LaPadula policy in its original formulation, and then we assess the challenges of adapting it to a distributed setting. Section \ref{sec:BelnapSynopsis} gives a brief review of the Logic used for dealing with the combination of policies. In Section \ref{sec:AspKB-MAC} we present our extended framework, and show how to precisely capture the Bell-LaPadula policy. We also discuss why the resulting framework is more flexible than the existing one. In Section \ref{sec:Conclusion} we conclude.

\begin{subsection}{Limitations of looking to the future}
\label{subsec:Limitations}
The framework we shall be dealing with throughout this paper is the formal language \textbf{AspectKB}. In that framework, which follows a process algebraic approach, the processes are modelled as actions taking place in specific locations, and interacting with other locations modelled as well. Furthermore, security policies can also be modelled, following an aspect-oriented manner. The policies can express their intentions by analysing the continuation (namely the process after the current action) of the involved processes, so that it is possible to know in advance what a process might do in the future. This reflects an information-flow style of providing security.

However, this information-flow style is not as adequate as it was for sequential programs. Indeed, since the only process that can be statically analysed is the one that continues after the current action, all the possible outcomes that may occur due to other processes could not be predicted. This means that, when deciding whether to allow the interaction to happen or not, it is necessary to look to the future of just one process, and this can lead to two possible ways of obtaining imprecise decisions, either \emph{over-approximation} or \emph{under-approximation}.

For understanding what over-approximation is, let us assume we \emph{pessimistically} expect that a particular action done by a process could, because of other processes we do not know about, lead to an insecure state. In this situation we may disallow the process to execute that action, but in some cases there might be no other process performing anything that would lead to an insecure state.

For understanding what under-approximation is, let us assume we \emph{optimistically} expect that a particular action done by a process will not lead to an insecure state because the very same process will not perform another related action that leads to such a state. In this situation we may allow the process to execute that action, but in some cases there might be some other process that makes the system reach some insecure state, due to some interactions that could have been avoided if the action was disallowed.

Let us discuss a simple example, without going into syntactic and semantic details, but still thinking about distributed processes and policies.

Let us think about a security policy where we have different security levels, and every location is assigned to some level. We do not want any information to be leaked from any security level to lower ones. Then, we should allow a process, running in a given location, to read data from another location, as long as the following two conditions are met: first, the other location, where the data is right now, is in a security level not higher than the one where the process is running; second, the process will not try, in the future, to write information to locations with security levels lower than the level of the location where the data is right now, since this writing may be influenced by the reading previously done.

Let us assume now a particular situation where we have 4 locations (say $A, B, C \text{ and } D$), and 3 security levels (say 1, 2 and 3). Let us assume the security levels are ordered as their values in natural numbers ($3 > 2 > 1$). Let us assume that location $A$ is in security level 1, locations $B$ and $C$ are both in security level 2, and location $D$ is in security level 3. Figure \ref{fig:examples} contains three cases of such a situation, showing the locations and their security levels in different layers.

Illustrated in Figure \ref{fig:examples:a}, there is a process in location $D$, that tries to read information from location $B$ at t1, and then tries to write some information to location $A$ at t2. This process should clearly be forbidden, because it does not meet the second condition of the policy we are trying to capture (although it meets the first one). Of course this can be done following the information-flow approach, looking to the future at t1, since we know that the process trying to read from $B$ will try to write to $A$, and this should not be allowed.

However, let us think about another case, illustrated in Figure \ref{fig:examples:b}. Let us say that the process running in location $D$ whose first action is to read information from location $B$ at t1 then tries to write some information to location $C$ (in the same level as $B$) at t2. This \emph{does} meet the second condition of the policy, since the only information the process could write to $C$ is what it has read from $B$. Therefore, this should be allowed.

But let us consider the next extension to the example, illustrated in Figure \ref{fig:examples:c}. Assume there is a fifth location $E$ that is in security level 3. Assume there is a process running in $E$ that writes some information to $D$ at t2, after the process running in $D$ has read from $B$ at t1. In this case, the future writing to $C$ by the process running in $D$ (which in this case will be done at t3) should be forbidden, because it might be influenced by the new information learned by location $D$ at t2. Anyway, since the process that writes to $C$ is not the same as the one running in $D$, the process algebraic way of modelling does not permit us to know in advance (at t1) that this will happen. If we had taken an approach of looking to the past, then we would have checked the insecure operation of writing to $C$ right in the moment of the writing (at t3), and we would have known that some information from $E$ would be leaked, and therefore we would avoid the write operation.

We could take the information-flow approach using over-approximation, and always avoid this type of write operation (e.g. from $D$ to $C$, since the former is in level 3 while the latter in level 2), but that would be very imprecise (and restrictive), since sometimes there is nothing insecure in doing this write operation, as shown in the case of Figure \ref{fig:examples:b}. Taking the information-flow approach using under-approximation would mean allowing the process in $D$ to perform the read and the subsequent write, since that write operation is not insecure. This will be secure enough in the case of Figure \ref{fig:examples:b}, but not in the case of Figure \ref{fig:examples:c}.

\begin{figure}[t]
  \centering
  \subfloat[Insecure case, detectable by information-flow approach.]{\label{fig:examples:a}\includegraphics[width=0.3\textwidth]{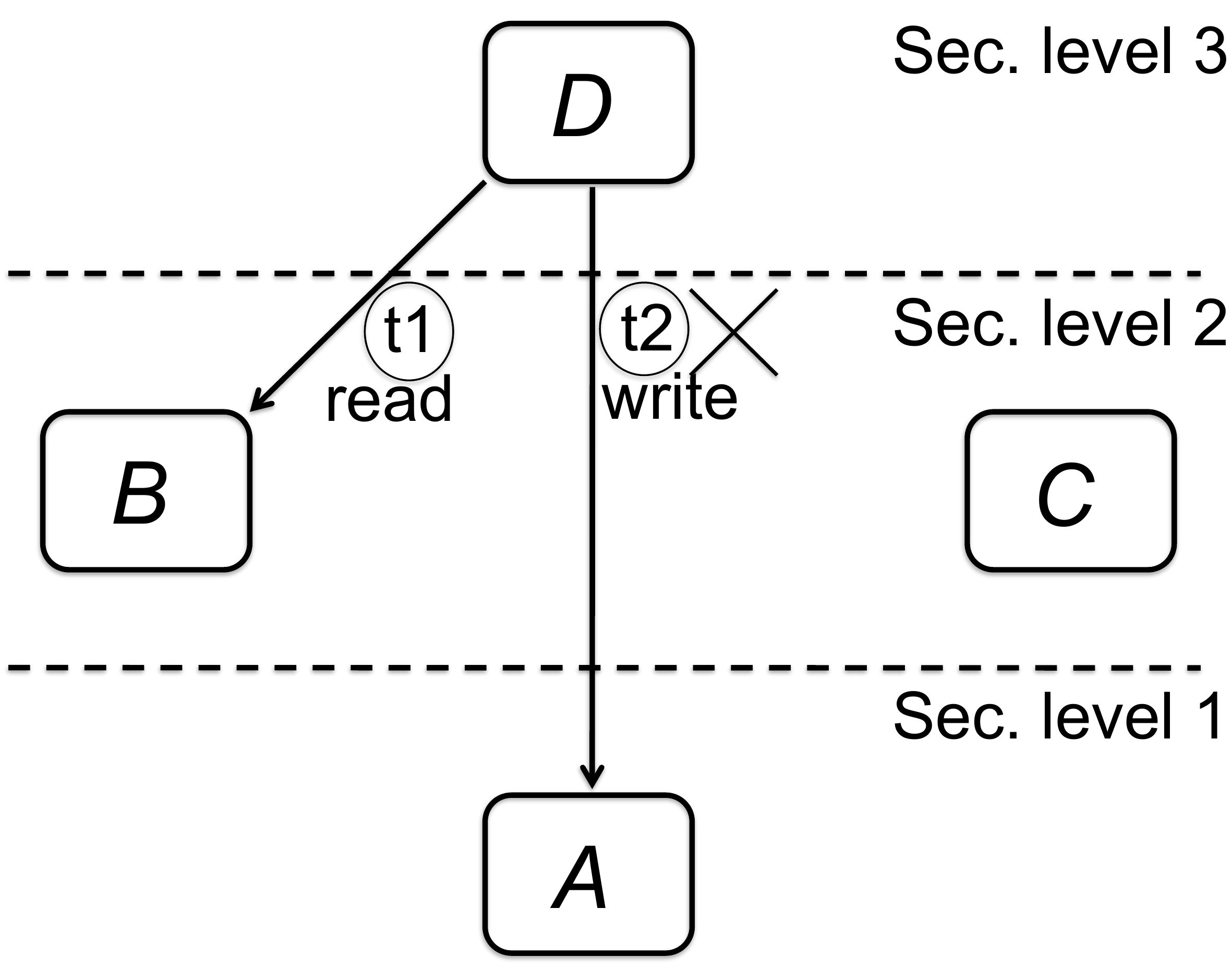}}
  \hspace{0.1cm}
  \vline
  \hspace{0.1cm}
  \subfloat[Secure case, but over-approximation would incorrectly disallow it.]{\label{fig:examples:b}\includegraphics[width=0.3\textwidth]{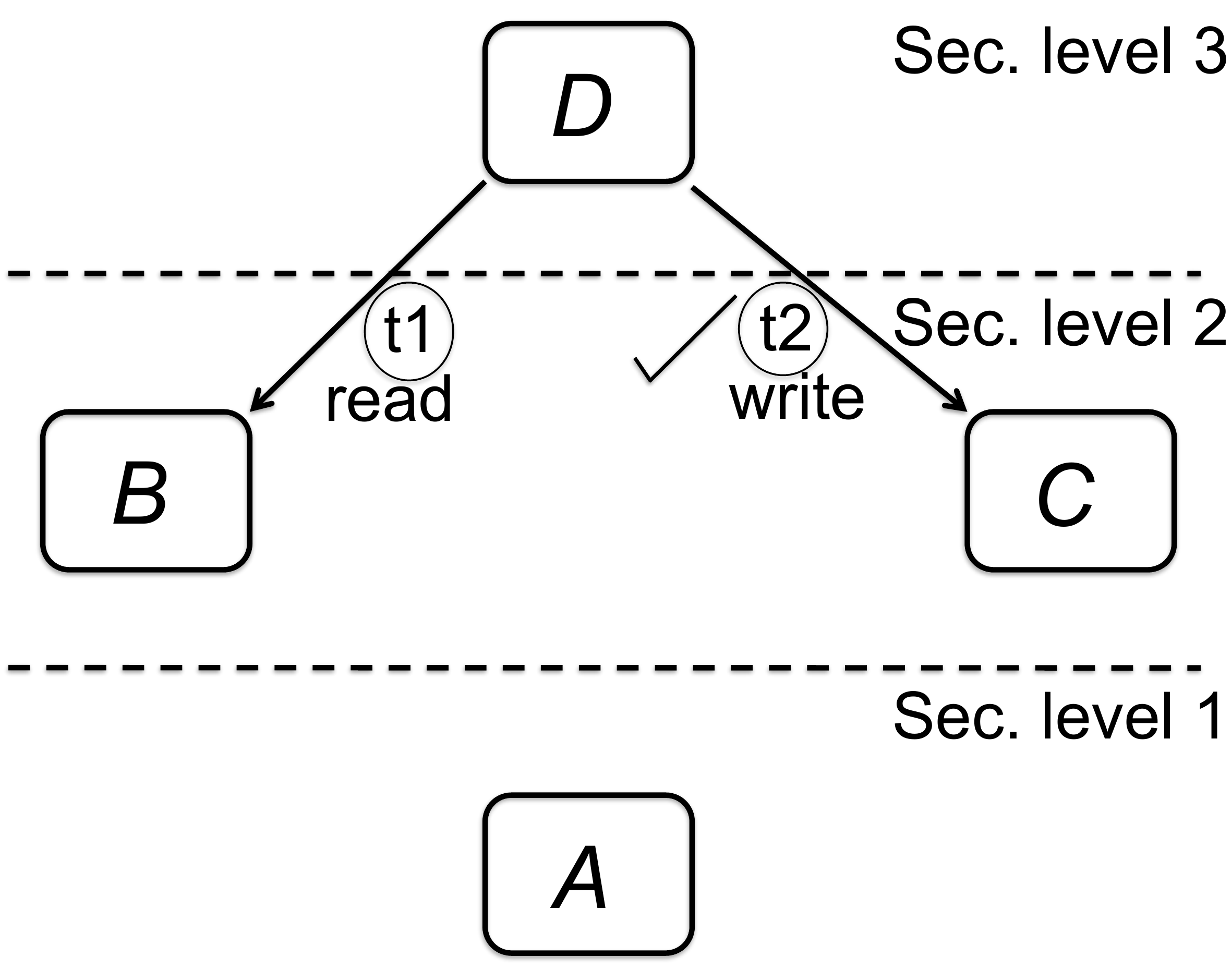}}
  \hspace{0.1cm}
  \vline
  \hspace{0.1cm}
  \subfloat[Insecure case, but under-approximation would incorrectly allow it.]{\label{fig:examples:c}\includegraphics[width=0.3\textwidth]{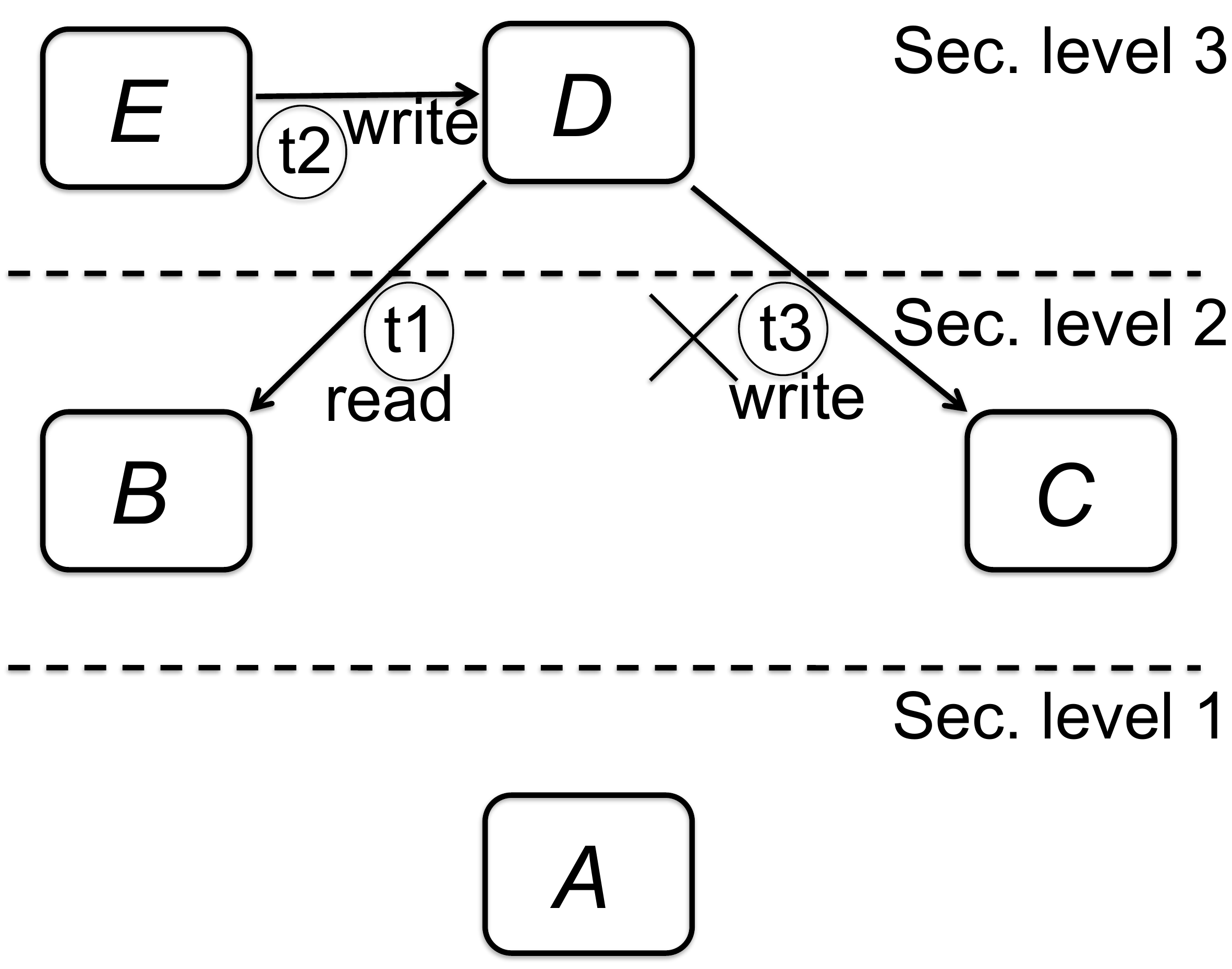}}
  \caption{Examples of situations that might happen.}
\label{fig:examples}
\end{figure}

So, we have found some possible situations where using an information-flow approach in a distributed setting is not completely precise, and therefore another approach might be taken, for instance looking to the past. In the rest of this work, we shall be studying how to deal with looking to the past, and how to extend our distributed-systems framework to achieve this. We shall see that the resulting framework allows us to combine both approaches, therefore obtaining the advantages of both of them. In particular, we shall see that the simple example we have seen is just one possible instance of something that can easily (and more precisely) be captured by the Bell-LaPadula policy.
\end{subsection}
\end{section}

\begin{section}{Assessment of the Bell-LaPadula model}
\label{sec:BLP}
In Subsection \ref{subsec:Limitations} we saw that for distributed systems the information-flow approach is not as adequate as it was for sequential programs. In this Section, we review another approach, the Bell-LaPadula (BLP for short) policy, and discuss the challenges of using it in a distributed setting, but aiming to show that this \emph{can be} as adequate as in its original formulation.

\begin{subsection}{The Operating System view of BLP}
\label{subsec:BLPorig}
The BLP model is the most traditional Mandatory Access Control model. Here we briefly introduce it, inspired by \cite{Dieter}, abstracting some unnecessary details that do not contribute to our study.
\begin{paragraph}{State.}
The computer system will be checked for security by looking into its state. For representing it, some sets must be introduced:
\begin{itemize}
\item $S$ is the set of subjects that may use the information stored in the system,
\item $O$ is the set of objects (pieces of information) stored in the system,
\item $A = \{\texttt{read}, \texttt{write}\}$ is the set of operations a subject may do over an object,
\item $L$ is a lattice of security levels.
\end{itemize}
Every state of the system is composed of a set of tuples of the form $(s, o, a)$ (each tuple would mean that subject $s$ is doing an $a$ operation over object $o$), and of a tuple of functions $(f_S, f_C, f_O)$ with types $S \rightarrow L$, $S \rightarrow L$ and $O \rightarrow L$. The functions are supposed to be total functions, and they will give, respectively, the maximum security level a subject can have (its \emph{clearance}), the \emph{current} security level a subject has\footnote{A subject can log into the system with a lower security level than its corresponding clearance. Once it did so, that security level cannot be changed until it logs in again.}, and the security level an object has (its \emph{classification}). Formally, a state $(B, F) \in \mathcal{B} \times \mathcal{F}$, where $F = (f_S, f_C, f_O)$, and where:
\begin{itemize}
\item $\mathcal{B} = \mathbb{P}(S \times O \times A)$
\item $\mathcal{F} = (S \rightarrow L)\times(S \rightarrow L)\times(O \rightarrow L)$
\end{itemize}
\end{paragraph}

\begin{paragraph}{Policies.}
The BLP model specifies two properties that every state should meet in order to be considered secure.

\begin{itemize}
\item \textbf{ss-property\footnote{For ``simple security''.}.} A state $(B, F)$ satisfies this property iff $\forall (s, o, a) \in B: a = \texttt{read} \implies f_S(s) \geq f_O(o)$. This means that each object being read by a subject should be in a level not higher than the level the subject is able to reach, which is usually called \emph{no read-up}.

\item \textbf{$\star$-property\footnote{Read ``\emph{star} property''. In some formulations of the BLP model, this property only consists of the first part because the ss-property uses $f_C$ instead of $f_S$, and then the second part is just a consequence. However, that kind of formulation is again too restrictive, since a subject cannot perform read operations in levels up to its clearance, but just up to the level it has logged in.}.} This property consists of two parts. A state $(B, F)$ satisfies the first part (let us name it \textbf{$\star$-property.1}) of this property iff $\forall (s, o, a) \in B: a = \texttt{write} \implies f_O(o) \geq f_C(s)$. This means that each object being written by a subject should be in a level not lower than the level the subject is currently in, which is usually called \emph{no write-down}. On the other side, a state $(B, F)$ satisfies the second part (let us name it \textbf{$\star$-property.2}) of this property iff $\forall (s, o, a) \in B: a = \texttt{write} \implies [\forall (s, o', a') \in B: a' = \texttt{read} \implies f_O(o) \geq f_O(o')]$. This means that if a specific subject (note the use of the same $s$ in both quantifications)
is operating with many objects, some being read and some being written, then no object being read could be in a higher level than any object being written. This prevents the subject to read some high-level object and then write a low-level one.
\end{itemize}
\end{paragraph}

A state is said to be $secure$ if it satisfies both properties.
\end{subsection}

\begin{subsection}{The challenges of distribution}
The BLP model was originally meant for Operating Systems. These have a particular feature: they are centralised, meaning that a central controller (i.e.~the Operating System) takes care of everything that happens in the system. In particular it can control (and in some cases restrict) the processes that try to access resources. Moreover, one key concept needed for checking BLP policy compliance is the \emph{state}, and since Operating Systems have a centralised state, they can do the calculations for knowing whether the BLP policy is met or not.

\begin{paragraph}{Lack of central controller.}
In a distributed setting we do not have any central controller; many locations run in parallel and share information, but no location can know what other locations are doing. Therefore, once a location is allowed access to some resource, there is no way other locations can forbid it from doing whatever it wants with the resource. In particular, there is no notion of state, processes interact and synchronise, but no central entity knows what has happened in the whole system so far.
\end{paragraph}

\begin{paragraph}{}
It should be clear that a distributed framework is not trivially able to meet security properties that were originally developed for simpler systems, as for example centralised ones or sequential programs. In the case of Information-Flow approach, we have seen some simple examples where we can lose precision. In the case of BLP, in the next Subsection we propose an extension that will help us to adapt the policy to a distributed setting.
\end{paragraph}
\end{subsection}

\begin{subsection}{Extending BLP}
\label{subsec:ExtendingBLP}
The original formulation of the BLP policy relies on three functions, two of which can be applied to every subject and one to every object. They can be computed by the Operating System every time an action is to be executed, to check whether the resulting state will still be secure, and then decide whether to allow the action or not. Here we propose an extension to their domains to have common signatures, since in a setting without a central controller we might want to call any of them with any possible entity of the system without distinguishing between objects and subjects. We also propose a fourth function which captures information about the past interactions for each entity. Later in the paper we will see that this latter function can be used to have a form of localised state.

The types of the three existing functions are then changed to $S\ \cup\ O \rightarrow L$ for all of them, and their definitions are extended in a straighforward way as follows:
$$
\forall o \in O, s \in S : f_S(o) = f_O(o) \land f_C(o) = f_O(o) \land f_O(s) = f_S(s)
$$

We call the new function $f_H$ since it keeps track of (a part of) the \emph{history} of the system. When we apply this function to a particular input subject (resp. object) we should learn what kinds of interactions the subject (resp. object) has been involved in during the past, therefore the output of the function would be a kind of current state of the argument subject (resp. object). To capture this notion of state, the function will not be fixed once and for all, as the original three functions were. Indeed, the output of this function will be:
\begin{itemize}
\item For a particular subject: the least upper bound of the security levels of all the objects read by the subject so far.
\item For a particular object: the least upper bound of the security levels of all the subjects that have written to the object so far.
\end{itemize}
Formally, this can be expressed as follows (assuming $(B, (f_S, f_C, f_O, f_H))$ to be some \emph{``virtual'' global} state that depends on the interactions that have happened and $(B', (f_S, f_C, f_O, f'_H))$ the next one):
$$
\begin{array}{lllll}
\forall (s, o, a) \in B: & ( & \,\,\,\,(\,\,\,(a = \texttt{read} \implies f_H(s) \geq f_O(o)) \,\,\, \land & & \\
& & \,\,\,\,\,\,\,\,\ (a = \texttt{write} \implies f_H(o) \geq f_C(s)) \,\,\,) & & \land \\
& & \multicolumn{2}{l}{\,\,\,\,\ \forall (s', o', a') \in B': (\, f'_H(o') \geq f_H(o) \land f'_H(s') \geq f_H(s))} & \,\,\,\,) \\
\end{array}
$$
This means that every time an interaction takes place, changing the state from $(B, (f_S, f_C, f_O, f_H))$ to some $(B', (f_S, f_C, f_O, f'_H))$, the output of $f'_H$ for some input may be higher than or equal to that of $f_H$. It can actually be higher depending on the values of the entities read/written, for keeping the resulting least upper bound we expect to have. Indeed, a very simple result tells us that for every set $\mathcal{L}$:
\begin{equation}
\label{lublast}
\sqcup (\mathcal{L}) = \sqcup \{ \sqcup (\mathcal{L} \setminus \{a\}) , a \} \qquad (\forall a \in \mathcal{L})
\end{equation}
And we should also observe that $\sqcup (\emptyset) = \bot$.

We shall use these four functions to capture this extended version of BLP in a distributed setting.
\end{subsection}

\end{section}

\begin{section}{A brief review of Belnap Logic}
\label{sec:BelnapSynopsis}
For granting access according to some security policy, the traditional boolean values (\TrueB\ and \FalseB) are enough: \TrueB\ grants while \FalseB\ denies access. However, for a distributed setting, where policies might be contradictory (or not sufficiently informative), those two values might not be enough. We shall consider an extension to the Boolean Logic proposed by Belnap~\cite{BelnapOrig}, which has been used for combining security policies~\cite{HuthBelnap}.

In this extension to the boolean logic, two more values are considered: $\bot$ and $\top$ (read ``bottom'' and ``top''). The traditional \TrueB\ would mean ``the policy accepts the interaction'' whereas the traditional \FalseB\  would mean ``the policy does not accept the interaction''. Since different locations might aim at different security properties, their policies could be contradictory or they may lack information about some particular interaction. These situations can be represented by the two extra values that we have: $\bot$ meaning ``no decision'' and $\top$ meaning ``contradiction'' or ``conflict''.

With this set of values, which we will call here \textbf{Four} (i.e.~\textbf{Four} = \{$\bot, \TrueB, \FalseB, \top$\}), it is possible to extend the usual boolean operations ($\land$ and $\lor$) and to define new ones ($\otimes$ and $\oplus$). For obtaining that, the set \textbf{Four} is equipped with two partial orderings, say $\leq_k$ and $\leq_t$, as shown in Figure \ref{fig:BelnapFig}.

The usual boolean $\land$ is extended as computing the greatest lower bound in the $\leq_t$ lattice, and the usual $\lor$ as computing the least upper bound (thereby obtaining the same results as in boolean logic if the operands belong to \{\TrueB, \FalseB\}). Analogously, the new operators over \textbf{Four} can be defined as computing the greatest lower bound (the $\otimes$ operator) and the least upper bound (the $\oplus$ operator), both in the $\leq_k$ lattice.\footnote{Notice that this could also be done by just extending the ``truth tables'' of the usual boolean operators and defining new ones for the new operators. That would mean, however, having not just 2 truth tables with 4 cells each (as in Boolean Logic) but 4 truth tables with 16 cells each, making it difficult to remember what each operator produces.}

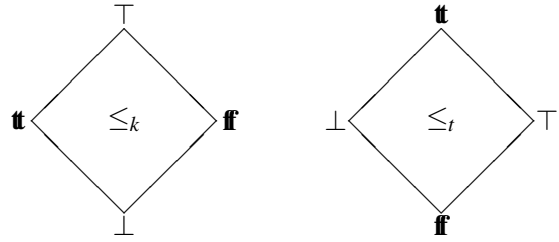
\begin{figure}[t]
\begin{center}
\begin{picture}(230,110)
\linethickness{1pt}
\put(40,90){\makebox(20,15){$\top$}}
\put(0,50){\makebox(20,15){\TrueB}}
\put(80,50){\makebox(20,15){\FalseB}}
\put(40,10){\makebox(20,15){$\bot$}}
\put(40,50){\makebox(20,15){$\leq_k$}}
\put(50,22){\line(1,1){35}}
\put(50,22){\line(-1,1){35}}
\put(50,92){\line(1,-1){35}}
\put(50,92){\line(-1,-1){35}}
\put(160,90){\makebox(20,15){\TrueB}}
\put(120,50){\makebox(20,15){$\bot$}}
\put(200,50){\makebox(20,15){$\top$}}
\put(160,10){\makebox(20,15){\FalseB}}
\put(160,50){\makebox(20,15){$\leq_t$}}
\put(170,22){\line(1,1){35}}
\put(170,22){\line(-1,1){35}}
\put(170,92){\line(1,-1){35}}
\put(170,92){\line(-1,-1){35}}
\end{picture}
\end{center}
\caption{The Belnap bilattice {\bf Four}: $\leq_k$ and $\leq_t$.} \label{fig:BelnapFig}
\end{figure}

The negation operator $\lnot$ is extended by leaving the two new values unchanged (i.e.~$\lnot \bot = \bot$ and $\lnot \top = \top$), and the implication $\Rightarrow$ is extended as follows:
$$
\begin{array}{lr}
p_1 \Rightarrow p_2 =
\left\{
\begin{array}{ll}
p_2 & \text{if } p_1 \leq_k \TrueB \\
\TrueB & \text{otherwise}
\end{array}
\right.
&
\forall p_1, p_2 \in \textbf{Four}
\end{array}
$$
Another useful operator is the priority $>$, which returns the first operand unless it is $\bot$, in which case it returns the second operand. This would always consider what the first operand suggests unless it has no decision, in which case the second operand is considered.
\end{section}

\begin{section}{Aspect-oriented framework for security}
\label{sec:AspKB-MAC}
As mentioned, the \textbf{AspectKB}~\cite{AspectKB} framework allows us to express location-based systems in a process-calculus-oriented manner. This is achieved by extending the KLAIM~\cite{KLAIM} coordination~\cite{Coordination} language. These \emph{located processes} interact with other locations when they try to gather (or put) information from (or into) them (maybe themselves), which are usually named \emph{tuple spaces}. The possibility of attaching to each location (regardless of whether it is a process or a tuple location) some security policy, which will govern the interactions the location may be involved in, turns the \textbf{AspectKB} language into an aspect-oriented language. Then, whenever an interaction takes place, the relevant policies are considered by the semantics to either grant or deny the interaction, using the four-valued Belnap Logic for deciding in a consistent way.

In this Section, an extension to that framework is made, mixing all process locations and tuple locations into just \emph{entity locations}, and attaching to them more aspects than just the security policies. The extra information attached to each location refers to security levels in the sense of a multilevel security policy. Moreover, the mechanisms of the language explicitly keep track of some information (at a certain level of abstraction) regarding the interactions that have taken place, giving the flavour of a \emph{localised state}, which the semantics of the language keep updated\footnote{As one can argue, having information inside the locations, namely the tuples, also gives us the flavour of state, yet that is information that changes according to what processes do, so we cannot rely on that information for guaranteeing any property.}.

Besides this, one can write Aspects using the extra information, which is basically the output of the functions mentioned in Subsection \ref{subsec:ExtendingBLP} (considering that every entity location can be either a subject and/or an object in the whole system, so every location can be a potential input to all those functions). Then, this will allow us to capture, among others, the BLP policies without losing precision.

Following this informal introduction to our extension, which we shall call \textbf{AspectKB+} due to its enhanced features, we shall present its formalities.

\begin{subsection}{Syntax and Semantics}
\label{subsec:syntaxsemantics}
\begin{paragraph}{Syntax.}
The syntax of the \textbf{AspectKB+} language is given in Tables \ref{tab:klaim:syn_NetProcess_wMAC} and \ref{tab:syn_Aspects_wMAC}. Table \ref{tab:klaim:syn_NetProcess_wMAC} gives the syntax for nets, the basic modules that can be described in the language. A net is a parallel composition of located processes and/or located tuples (data), together with an annotation explained below. Each process can be a parallel composition of processes, a non-deterministic choice between processes following an action, or a replicated process. A process not performing any action shall be written \textbf{0}. The allowed actions are reading from a location (\textbf{in} and \textbf{read}, resp. with or without deleting the data read) or writing to it (\textbf{out}).

Every location has an annotation $\w$, whose first part ($lst$) is intended to keep track of the interactions that the location has been involved in. This \emph{localised state} consists of 4 pieces of information (namely $\gamma^S, \gamma^C, \gamma^H, \text{ and } \gamma^O$) that are elements of the lattice $L$ of security levels (introduced in Subsection \ref{subsec:BLPorig}). Since every location can be input to the four functions $f_S, f_C, f_H \text{ and } f_O$, and since the result of evaluating them is in $L$, we can keep attached to each location the result of evaluating each of those four functions.


\begin{table}[t]
$$
\begin{array}{l@{\quad}rcl}
N  \in  {\bf Net} &
N & ::= & N_1\netpar N_2\ \mid\ \Localised{l}{\w}{P} \mid\
\Localised{l}{\w}{\val{\veck{\YZl}}}
\\
P  \in  {\bf Proc} &
P & ::= &  P_1\ppar P_2\ \mid\ \sum_i a_i.P_i\ \mid\ *P
\\[.5ex]
a  \in  {\bf Act} &
a & ::= & \locate{\OutM{\veck{\ell}}}{\ell}\ \mid\ \locate{\InM{\veck{\ell^{\lambda}}}}{\ell} \mid\
\locate{\ReadM{\veck{\ell^\lambda}}}{\ell}
\\
\Ln,\Lnt \ \in \ {\bf Loc} &
\Ln & ::= & u \  \mid\ \YZl \hfill 
  \Lnt\  ::= \  \Ln \mid\ !u  \\
  \w  \in  {\bf Annot} &
\w & ::= & < lst, pol > \\
  lst  \in  {\bf LocSt} &
lst & ::= & < \gamma^S, \gamma^C, \gamma^H, \gamma^O > \\
\gamma \in L & \multicolumn{3}{l}{left\text{ }implicit}
\end{array}
$$
\caption{{\bf AspectKB+} Syntax -- Nets, Processes, Actions and States.} \label{tab:klaim:syn_NetProcess_wMAC}
\end{table}

\begin{table*}[t]
$$
\begin{array}{lrcl}
pol \in {\bf Pol} & 
pol & ::= & 
asp \mid
\neg pol \mid 
pol \oplus pol \mid 
pol \otimes pol \mid pol \Rightarrow pol \mid \\
& & &
pol > pol \mid pol \wedge pol \mid pol \vee pol \mid \TrueS \mid \FalseS \\
asp  \in  {\bf Asp} &
asp & ::= & [rec\ \IF\ cut : cond ]
\\
cut  \in  {\bf Cut} &
cut & ::= & \Lc :: a^t\,.\, X
\\
a^t \in {\bf Act}^t &
a^t & ::= & \locate{\OutM{\veck{\ell^t}}}{\ell}\ \mid\ \locate{\InM{\veck{\ell^{t\lambda}}}}{\ell}\ \mid\
\locate{\ReadM{\veck{\ell^{t\lambda}}}}{\ell}
\\
rec \in {\bf Rec} &
rec & ::= &
\ell_1 = \ell_2 \mid
\neg rec \mid 
rec  \oplus  rec  \mid
rec  \otimes  rec \mid
rec \wedge rec \mid
rec \vee rec \mid
\\
&&&
rec \Rightarrow rec \mid
\TrueS \mid
\FalseS \mid
a\ \hbox{\bf occurs-in}\ X \mid
v_1 \geq v_2
\\
cond \in {\bf Cond} &
cond & ::= &
\ell_1 = \ell_2 \mid
\neg cond \mid 
cond_1  \land  cond_2  \mid
cond_1  \lor  cond_2 \mid
\\
&&&
\TrueS \mid
\FalseS \mid\ 
a\ \hbox{\bf occurs-in}\ X\\ 
v \in {\bf Lev} & v & ::= &
S_s \mid C_s \mid H_s \mid O_t \mid H_t \mid \gamma
\\
& \ell ^t & ::= & \ell \mid \_
\hfill 
  \ell^{t\lambda}\  ::= \  \Ln \mid\ \_ \hfill
\end{array}
$$
\caption{{\bf AspectKB+} Syntax - Aspects for Security Policies.} \label{tab:syn_Aspects_wMAC}
\end{table*}

The second part ($pol$) of the annotation in every location is the actual security policy governing the location, which has to be expressed using the syntax of Table \ref{tab:syn_Aspects_wMAC}. The policy can be a Belnap combination of policies, a boolean value, or a single aspect. This latter consists of a $cut$ (the action, together with its continuation, to be trapped by the aspect), a condition $cond$ (a boolean applicability condition) and a recommendation $rec$ (a four-valued Belnap Logic advice for the aspect). To define an aspect, one may refer to the security levels stored in the trapped interaction or to a single value from the lattice $L$. To do the former, one can write an aspect naming some of the five syntactic names ($S_s, C_s, H_s, O_t, H_t$) specified in the category $v \in \textbf{Lev}$, which will later be matched by the semantics to the specific values kept in the trapped interaction. To do the latter, one can write an aspect providing a specific value from $L$, as the category $v \in \textbf{Lev}$ permits (by having $\gamma$ among its choices). Finally, the \textbf{occurs-in} operator, which can be easily defined in a compositional way, checks whether the action occurs in the continuation process.
\end{paragraph}

\begin{paragraph}{Semantics.}
The semantics is given by a one-step reduction relation. It makes use of a structural congruence on nets (defined in Table \ref{tab:klaim:sem_structural_congruence_wMAC}), and also of an operation $match$, for matching input patterns to actual data, which could easily be defined in an inductive way by the structure of its arguments.


\begin{table*}[t]
$$
\begin{array}{l}
\multicolumn{1}{c}{
\Inference{ N_1 \rightarrow N_1'}{ N_1 \netpar
N_2 \rightarrow  N_1'\netpar N_2}
\qquad\qquad\qquad
\Inference{N \equiv M \quad M \rightarrow M' \quad M' \equiv
N'}{ N \rightarrow N'}
}
\\[3.5ex]
(\LocalisedSubject{\w_s}{\mathbf{read}(\veck\Lat)@l_t.P+ \cdots}) \netpar
(\LocalisedTarget{\w_t}{\langle \veck\YZl \rangle})
\\
\qquad \begin{array}{cl}
\rightarrow &
\left\{
\begin{array}{ll}
\LocalisedSubject{\w'_s}{P\theta} \netpar \LocalisedTarget{\w_t}{\langle \veck\YZl \rangle} & \textrm{if } b\ \land \ match(\veck\Lat;\veck\YZl)= \theta \\
\LocalisedSubject{\w_s}{\nil} \netpar
\LocalisedTarget{\w_t}{\langle \veck\YZl \rangle} & \textrm{if } \neg b
\end{array}
\right.\\[2ex]
\multicolumn{2}{l}{\textrm{where } \w_{\delta} = < < \gamma^S_{\delta}, \gamma^C_{\delta}, \gamma^H_{\delta}, \gamma^O_{\delta} > , pol_{\delta}>, \qquad (\delta \in \{s, t\});}
\\
\multicolumn{2}{l}{\textrm{and where } b = \grant{\db{pol_s \oplus pol_t}(l_s::\mathbf{read}(\veck\Lat)@l_t.P{, 
< \gamma^S_s, \gamma^C_s, \gamma^O_t, \gamma^H_s, \gamma^H_t >
})};}
\\
\multicolumn{2}{l}{\textrm{and where } \w'_s = < < \gamma^S_s, \gamma^C_s, (\gamma^H_s \sqcup (\gamma^O_t \sqcup \gamma^H_t)), \gamma^O_s > , pol_s >.}
\end{array}
\\[7.5ex]
(\LocalisedSubject{\w_s}{\mathbf{in}(\veck\Lat)@l_t.P+ \cdots}) \netpar
(\LocalisedTarget{\w_t}{\langle \veck\YZl \rangle})
\\
\qquad \begin{array}{cl}
\rightarrow &
\left\{
\begin{array}{ll}
\LocalisedSubject{\w'_s}{P\theta} & \textrm{if } b\ \land \ match(\veck\Lat;\veck\YZl)= \theta \\
\LocalisedSubject{\w_s}{\nil} \netpar
\LocalisedTarget{\w_t}{\langle \veck\YZl \rangle} & \textrm{if } \neg b
\end{array}
\right.\\[2ex]
\multicolumn{2}{l}{\textrm{where } \w_{\delta} = < < \gamma^S_{\delta}, \gamma^C_{\delta}, \gamma^H_{\delta}, \gamma^O_{\delta} > , pol_{\delta}>, \qquad (\delta \in \{s, t\});}
\\
\multicolumn{2}{l}{\textrm{and where } b = \grant{\db{pol_s \oplus pol_t}(l_s::\mathbf{in}(\veck\Lat)@l_t.P{, 
< \gamma^S_s, \gamma^C_s, \gamma^O_t, \gamma^H_s, \gamma^H_t >
})};}
\\
\multicolumn{2}{l}{\textrm{and where } \w'_s = < < \gamma^S_s, \gamma^C_s, (\gamma^H_s \sqcup (\gamma^O_t \sqcup \gamma^H_t)), \gamma^O_s > , pol_s >.}
\end{array}
\\[7.5ex]
(\LocalisedSubject{\w_s}{\mathbf{out}(\veck\YZl)@l_t.P+\cdots})
\netpar (\LocalisedTarget{\w_t}{Q})
\\
\qquad \begin{array}{cl}
\rightarrow &
\left\{ 
\begin{array}{ll}
\LocalisedSubject{\w_s}{P} \netpar \LocalisedTarget{\w'_t}{\langle \veck\YZl \rangle}
  \netpar \LocalisedTarget{\w_t}{Q}  & \textrm{if } b\\
\LocalisedSubject{\w_s}{\nil} \netpar \LocalisedTarget{\w_t}{Q}  & \textrm{if } \neg b
\end{array}
\right.\\[2ex]
\multicolumn{2}{l}{\textrm{where } \w_{\delta} = < < \gamma^S_{\delta}, \gamma^C_{\delta}, \gamma^H_{\delta}, \gamma^O_{\delta} > , pol_{\delta}>, \qquad (\delta \in \{s, t\});}
\\
\multicolumn{2}{l}{\textrm{and where } b = \grant{\db{pol_s \oplus pol_t}(l_s::\mathbf{out}(\veck\YZl)@l_t.P{, 
< \gamma^S_s, \gamma^C_s, \gamma^O_t, \gamma^H_s, \gamma^H_t >
})};}
\\
\multicolumn{2}{l}{\textrm{and where } \w'_t = < < \gamma^S_t, \gamma^C_t, (\gamma^H_t \sqcup (\gamma^C_s \sqcup \gamma^H_s)), \gamma^O_t > , pol_t >.}
\end{array}
\end{array}
$$
\caption{Reaction Semantics of {\bf AspectKB+} .}
\label{tab:sem_two_wMAC}
\end{table*}

\begin{table}[t]
$$
\begin{array}{lr}
\begin{array}{rcl}
l::^{\w} P_1 \ppar P_2 &\equiv& l::^{\w}P_1 \netpar l::^{\w}P_2
\\[1ex]
l ::^{\w}\ * P  &\equiv& l ::^{\w}\ P \ppar\ * P
\\[1ex]
l ::^{\w}\ P &\equiv& l ::^{\w} P \netpar l ::^{\w} {\bf 0}
\end{array}
& \hspace{2cm}
\begin{array}{rcl}
\multicolumn{3}{c}{\Inference{N_1 \equiv N_2}{N \netpar N_1 \equiv N \netpar N_2}}
\end{array}
\end{array}
$$
\caption{Structural Congruence.}
\label{tab:klaim:sem_structural_congruence_wMAC}
\end{table}


The reaction rules (defined in Table \ref{tab:sem_two_wMAC}) prescribe how the system may evolve in the presence of some \emph{process location} and some \emph{target location}.

In the ``where''
lines of each rule, the boolean condition $b$ is obtained by evaluating the security policies of the locations involved in the computation using the evaluation function \doublebracketleft.\doublebracketright\ (formally defined in Subsection \ref{subsec:meaning_pol}). This is done to either allow or disallow the process to compute, and for this it also makes use of the function {\sf grant} (also formally defined in Subsection \ref{subsec:meaning_pol}) for turning four-valued Belnap truth values into boolean truth values. If the action was disallowed the involved process simply terminates, thereby becoming just a \textbf{0}; otherwise the process evolves as the next paragraphs explain.

In the case of a \textbf{read} or \textbf{in} action, the process location $l_s$ is subject to a substitution, using the result of the matching done with the $match$ operation. Moreover, the localised state of that location might be modified, changing the historic component of its annotation by the least upper bound of the previous value and the security level of the target location $l_t$. This follows the suggestion of Equation (\ref{lublast}).

In the case of an \textbf{out} action, the data is stored in the target location. However, this is not done directly, but actually another ``virtual'' location is created, with a special localised state. This is intended to permit the virtual location holding the pre-existing process $Q$ to keep running as it was, without being interfered with. The virtual location now holding the data has an historic component on the annotation that is the least upper bound of the previous value in the location $l_t$ and the security level of the process location $l_s$ that has written the data. Of course it is possible that the value is the same as in the original $l_t$.

To simulate the log-in of a subject in a lower level than its clearance, a process can be annotated with a value for $\gamma^C$ lower that the $\gamma^S$. This value will then never change, just as the $\gamma^S$ and $\gamma^O$ components of the localised state. Note also that the security policy annotating each location never changes either.

\end{paragraph}
\end{subsection}

\begin{subsection}{Meaning of policies and granting access}
\label{subsec:meaning_pol}
In the ``where'' lines of each semantic rule there is a check that tells whether the interaction should be allowed. For this purpose, the policies of both locations taking part in the interaction are combined using the Belnap operator $\oplus$, and the result of the evaluation by the operator \doublebracketleft.\doublebracketright\ is passed to the function {\sf grant}.

The function {\sf grant} is defined by $\grant{p} = p \leq_k \TrueB$, for all $p$ in \textbf{Four}. The aim of granting access whenever the result is less than or equal to \TrueB\ is for doing so not only if both policies agree with this, but also if some of the policies lack some decision, because this would mean that it does not actually forbid the interaction. This is related to the use of $\oplus$ for combining the policies, and the aim is that whenever the policies are contradictory, the result of the evaluation by \doublebracketleft.\doublebracketright\ gives $\top \in \textbf{Four}$, thereby denying access as long as at least one policy has evidence that the interaction should be disallowed\footnote{This follows a conservative principle, as to actually grant access there should not be any policy at all denying the interaction.}.

The evaluation function \doublebracketleft.\doublebracketright\ (Table \ref{tab:aspectkb:policy_semantics_wMAC}) is defined inductively on the structure of the (infix) policy. The base cases are when the policy is just a constant (\TrueS\ or \FalseS) and when it is just an aspect (i.e.~it belongs to \textbf{Asp}). In this latter case, the first (postfix) parameter, a \emph{specific} action with continuation, is checked against the $cut$ of the aspect, a \emph{generic} action with continuation, using the function $check$, which could easily be defined in an inductive way by the structure of its arguments. This is achieved using a function $extract$, which produces the list of literals that occur in an action with continuation in a way that, for instance, $extract(\ell :: \mathbf{out}(\ell_1^t,\cdots,\ell_n^t)@\ell'.X) = [\ell,\mathbf{out},\ell_1^t,\cdots,\ell_n^t,\ell',X]$, which is done by just pattern matching the components of the given parameter and then pushing them into a list. The function $check$ determines whether there is a substitution $\theta$ that can be performed in the $cut$ that matches the parameter given to the \doublebracketleft.\doublebracketright. This is needed because the $cut$ can possibly consist of variables for representing the locations and even the arguments of the action in the $cut$ may not be specified. If there is such $\theta$, the $rec$ and the $cond$ are substituted using it to determine the result. This is achieved using the usual two-valued meaning $\dt{cond}$, which could be straightforwardly adapted to a four-valued meaning $\dt{rec}$.

Due to the semantics of Table \ref{tab:sem_two_wMAC}, the first parameter will always be the actual action taking place.


\begin{table*}[t]
$$
\begin{array}{l}
\db{[rec\ \IF\ cut: cond]}(l :: a\,.\, P{, < \gamma_S, \gamma_C, \gamma_O, \gamma_{Hs}, \gamma_{Ht} >})
\quad = \quad
\\
\qquad
\left(\begin{array}[c]{l}
\hbox{case $check(\,extract(cut)\,;\,extract(l :: a\,.\, P))$ of}\ \\ 
\qquad\qquad \begin{array}[t]{l@{\quad}l}
\Fail: & \bot \\
\theta: & 
\left\{
\begin{array}{ll}
\dt{(rec\ \theta){\theta'} } & \textrm{if } \dt{cond\ \theta}
  \\
\bot & \textrm{if } \neg \dt{cond\ \theta}
\end{array}
\right.
  \\
& {\text{where } \theta' = [\gamma_S / S_s, \gamma_C / C_s, \gamma_O / O_t, \gamma_{Hs} / H_s, \gamma_{Ht} / H_t]}
\end{array}
\end{array}\right)
\\
[6ex]
\db{\neg pol}(N{, \Gamma}) \quad = \quad \neg(\db{pol}(N{, \Gamma}))
\\
[0ex]
\db{pol_1\ \phi\ pol_2}(N{, \Gamma})
\quad = \quad
(\db{pol_1}(N{, \Gamma}))\ \phi\ (\db{pol_2}(N{, \Gamma})), \ (\phi \in \{\oplus,\otimes,\Rightarrow,>,\wedge,\vee\})
\\
[0ex]
\db{\TrueS}(N{, \Gamma}) \quad = \quad
\TrueB
\\
\db{\FalseS}(N{, \Gamma}) \quad = \quad
\FalseB
\end{array}
$$
\caption{Meaning of Policies in {\bf Pol} for {\bf AspectKB+}.}
\label{tab:aspectkb:policy_semantics_wMAC}
\end{table*}

The second (postfix) parameter (consisting of five values in the lattice $L$) is used to produce another special substitution ($\theta'$) that is also used (together with $\theta$) to determine the result of the recommendation $rec$. Due to the semantics of Table \ref{tab:sem_two_wMAC}, the security levels annotated in the actual interacting locations are given here. Indeed, those taken from the target location are the ones identifying the \emph{classification} ($\gamma^O_t$) of the location and the \emph{historic} annotation ($\gamma^H_t$). Those taken from the process location are the ones identifying the \emph{clearance} ($\gamma^S_s$) and the \emph{current level} ($\gamma^C_s$) of the location and the \emph{historic} annotation ($\gamma^H_s$).

It should be noticed that, while the $\theta$ substitutes according to some checking performed between the $cut$ and the first parameter (the actual action), the $\theta'$ substitutes according to the five syntactic names prescribed by the syntax of Table \ref{tab:syn_Aspects_wMAC}, in the $v \in \textbf{Lev}$ meta-variable. Therefore, when describing a system in \textbf{AspectKB+}, these syntactic names could be used to describe recommendations ($rec$) that will later be used to check actual security levels of the interacting locations, as already pointed in Subsection \ref{subsec:syntaxsemantics}.
\end{subsection}

\begin{subsection}{Capturing BLP in \textbf{AspectKB+}}
\label{subsec:CapturingBLP}
Having developed our formal framework, we shall show how the extended BLP policy of Subsection \ref{subsec:ExtendingBLP} can be elegantly captured. We shall also show that we can easily decide which cases of the example in Subsection \ref{subsec:Limitations} are secure and which are not, without losing any precision, unlike the information-flow approach.

Remember that \textbf{AspectKB+} is a process calculus, and even though in the original formulation of BLP the compliance of a state with the policy is checked in \emph{every} state, we can just check if a \emph{transition} might take us to an ``insecure state''. Also remember that \textbf{AspectKB+} provides us with the possibility, when describing aspects, of writing in the recommendation $rec$ the five syntactic names we have mentioned, which later will be substituted by the evaluation function \doublebracketleft.\doublebracketright. So basically using those distinctive names we aim to capture the BLP policy.

\begin{paragraph}{The first aspects.} Let us focus first on the \textbf{ss-property}, which prescribes that a subject cannot read an object that has higher security level than itself. The operations that can read information from other locations are the \textbf{read} and the \textbf{in} actions. So the aspects that capture the \textbf{ss-property} are the following:
\begin{equation}
\label{eq:ss-read}
\AspectEqbegin
S_s \geq O_t \textif l_s::\textbf{read}(-)@l_t.P : \TrueS
\AspectEqend[]
\end{equation}
\begin{equation}
\label{eq:ss-in}
\AspectEqbegin
S_s \geq O_t \textif l_s::\textbf{in}(-)@l_t.P : \TrueS
\AspectEqend[]
\end{equation}
Note that each aspect is trapping a particular operation, without caring about the parameters and with a trivial applicability condition. Whenever some of these aspects trap an action, the recommendation will be considered, granting access only if the security level of the subject is not lower than that of the object, since the two names $S_s$ and $O_t$ will then be replaced by the corresponding security levels of the actual interacting locations, thanks to Tables \ref{tab:sem_two_wMAC} and \ref{tab:aspectkb:policy_semantics_wMAC}.

For the \textbf{$\star$-property.1}, which prescribes that a subject cannot write any object that has lower security level than the level the subject is currently in, we have to follow a similar approach. Considering that the write operations are the \textbf{out} and the \textbf{in} (since deleting data is a form of write, because some implicit information could be communicated), the aspects are as follows:
\begin{equation}
\label{eq:star1-out}
\AspectEqbegin
O_t \geq C_s \textif l_s::\textbf{out}(-)@l_t.P : \TrueS
\AspectEqend[]
\end{equation}
\begin{equation}
\label{eq:star1-in}
\AspectEqbegin
O_t \geq C_s \textif l_s::\textbf{in}(-)@l_t.P : \TrueS
\AspectEqend[]
\end{equation}
Whenever some of these aspects trap an action, the recommendation will only grant access if the security level of the object is not lower than the one the subject is currently in (note the use of $C_s$ instead of $S_s$).
\end{paragraph}

\begin{paragraph}{The \textbf{$\star$-property.2}.}
Now let us consider the \textbf{$\star$-property.2}, which was basically the one that initiated the proposal made in this paper, due to the difficulty of capturing it precisely in a distributed setting. Note that the semantics of \textbf{AspectKB+} will keep track of the least upper bound of the security levels of the objects read by a particular subject location, because it updates it whenever the subject reads something that is not lower than the current value.  A similar observation can be done for the object locations.

Let us consider a subject location, which might have read some high information as long as its security level allows it (otherwise either aspect (\ref{eq:ss-read}) or (\ref{eq:ss-in}) would have denied it). Any subsequent write to a low location must be denied, and in principle either aspect (\ref{eq:star1-out}) or (\ref{eq:star1-in}) might decide this, unless the subject is logged into the system with a low security level. In any case, using the localised state that we have in the subject location, and making use of the $H_s$ syntactic name provided by the syntax for expressing aspects, we define the following aspects:
\begin{equation}
\label{eq:star2-subjectout}
\AspectEqbegin
O_t \geq H_s \textif l_s::\textbf{out}(-)@l_t.P : \TrueS
\AspectEqend[]
\end{equation}
\begin{equation}
\label{eq:star2-subjectin}
\AspectEqbegin
O_t \geq H_s \textif l_s::\textbf{in}(-)@l_t.P : \TrueS
\AspectEqend[]
\end{equation}
They can be understood in a very similar way as aspects (\ref{eq:star1-out}) and (\ref{eq:star1-in}), with the difference being that they are considering the localised state of the subject location, instead of the level where the subject has logged into the system.

Analogous considerations can be done for an object location, and we can define the following aspects for finishing to capture the whole BLP policy:
\begin{equation}
\label{eq:star2-targetread}
\AspectEqbegin
S_s \geq H_t \textif l_s::\textbf{read}(-)@l_t.P : \TrueS
\AspectEqend[]
\end{equation}
\begin{equation}
\label{eq:star2-targetin}
\AspectEqbegin
S_s \geq H_t \textif l_s::\textbf{in}(-)@l_t.P : \TrueS
\AspectEqend[]
\end{equation}
\end{paragraph}

\begin{paragraph}{Combining the aspects.}
After defining these eight aspects, the idea is to combine and attach them to every location, so every time an interaction is to take place, the semantics will consider all the aspects before allowing the interaction to happen.

Since the BLP model says that a state is secure if \emph{both} properties are satisfied, then we need to make sure that none of the aspects representing the properties detects a possible insecure interaction, as that would mean that \emph{at least one} of the properties is not satisfied. For capturing this situation, again the Belnap operator that must be used to combine the aspects for attaching them to the locations is $\oplus$.

Now we are ready to state our first Lemma:
\begin{lemma}
If a distributed system is insecure in the sense of Section \ref{subsec:ExtendingBLP}, then some of the aspects from (\ref{eq:ss-read}) to (\ref{eq:star2-targetin}) will deny the insecure interaction.
\end{lemma}
For the converse we need to make an extra observation, discussed in the following Sub-subsection.
\end{paragraph}

\begin{subsubsection}{Initialising the historic value}
\label{subsubsec:initialising}

 The aspects just defined will check, among other values, the historic component $\gamma^H$ attached to each location, and that value will be kept updated by the semantics. But, initially, one must give a particular value for the component. The chosen value will not affect the correctness of the aspects detecting insecure interactions, but to fulfil our requirement that we should not lose any precision while doing so (unlike the information-flow approach) the value should be $\bot \in L$. This follows the suggestion of Equation (\ref{lublast}) and the observation just after it.
Now we are ready to state our the converse of the previous Lemma:
\begin{lemma}
If some of the aspects from (\ref{eq:ss-read}) to (\ref{eq:star2-targetin}) deny an interaction, then the hypothetical resulting global state, if the interaction was actually allowed, is insecure in the sense of Section \ref{subsec:ExtendingBLP}.
\end{lemma}

Furthermore, we can now easily verify that the three examples of Figure \ref{fig:examples} are precisely captured. In particular, it should be taken into account what could happen after the process in location $E$ writes to location $D$ (Figure \ref{fig:examples:c}). For the process in $D$ to be actually influenced by this, it must explicitly read the data, since the semantics of {\textbf{AspectKB+}} will put it in another ``virtual'' location, with a higher historic component. So if the process is influenced, then at t3 the aspects (actually aspect (\ref{eq:star2-subjectout})) will prevent the write to $C$, otherwise the write will be allowed.
\end{subsubsection}
\end{subsection}

\begin{subsection}{A very simple example}
Let us now consider a very simple example to show how to combine looking to the future and to the past. Assume an airline has a database where information about the passengers is kept. The historic component of the database location is initialised to $\bot \in L$ so any process could read from it, but after some data is written, only some processes could do so, according to the security level of the data written. The aspect that prescribes this is:
\begin{equation}
\label{eq:airlinepast}
\AspectEqbegin
clearance_u \geq history_{AirlineDB} \\
\textif u::\textbf{read}(\texttt{pass},-)@AirlineDB.P : \TrueS
\AspectEqend[]
\end{equation}
As one can notice, this is a special case of aspect (\ref{eq:star2-targetread}), but it is written like this here to emphasise the example.

One of the process locations that will not be allowed to read data from the database due to the previous aspect is the Government, whose clearance should not be enough to satisfy the $rec$ of the aspect. Indeed, the historic component of the database should be high enough since the data written in there might be sensitive for the passengers.

However, in times of heightened security due to probable threats, the Government should be able to audit the passengers, therefore allowing it to read the database is necessary. Anyway, this should be allowed as long as the Government will not, later, give the passengers' data to the Press, to keep satisfying the right to privacy of the passengers. The following aspect prescribes this:
\begin{equation}
\label{eq:airlinefuture}
\AspectEqbegin
\lnot(\textbf{out}(data)@PressRelease \ \hbox{\bf occurs-in}\ P) \\
\textif Government::\textbf{read}(\texttt{pass},data)@AirlineDB.P : \textbf{test}(threatlevel,\texttt{high})@AirlineDB
\AspectEqend[]
\end{equation}
This is just a \emph{little} aspect that looks to the future\footnote{In \cite{AspectEHR} there are many more realistic examples that look to the future, in the Electronic Health Records domain.}, where we see how this is achieved. In the presence of this aspect, the Government will be allowed to perform the read action, as long as there is a tuple $< threatlevel,\texttt{high} >$ in the Airline database (i.e. the Airline was already notified of the heightened security situation), and also as long as the Government process trying to read the data will not leak the data to the Press in the future.

But one of the conditions is set in the $cond$ of the aspect whereas the other in the $rec$. The reason is related to the fact that this aspect is a temporary one, and the aim is to combine it with the previous one. Moreover, the combination should be done in a way that the Government should actually be allowed to read the database, although the pre-existing aspect (aspect (\ref{eq:airlinepast})) might deny this. Therefore, the operator needed for combining the two aspects is the priority $>$, and then the whole security policy for the Airline database would be (\ref{eq:airlinefuture}) $>$ (\ref{eq:airlinepast})\footnote{Note that using this policy with the priority, the aspect (\ref{eq:airlinefuture}) could even remain there, instead of just being a temporary one, since it will be ignored in most of the cases, as long as the tuple $< threatlevel,\texttt{high} >$ is removed after the situation is normalised.}.

With that, if the process location is the Government and the heightened security situation is declared, then aspect (\ref{eq:airlinefuture}) will be considered. Otherwise, either the action will not be trapped by the aspect (if the process location is not the Government) or the condition $cond$ will be \FalseB\ (if the threat level is not high), resulting in both cases in a $\bot \in \textbf{Four}$ for aspect (\ref{eq:airlinefuture}), considering then the aspect (\ref{eq:airlinepast}).

This example, even though it is very simple, clearly shows three features of our framework:
\begin{itemize}
\item The use of Aspects for security allows us to temporarily modify a distributed system without having to dig into the bussiness logic of the processes.
\item The use of the four-valued Belnap Logic allows us to easily combine policies, providing flexibility for the aspect-oriented framework.
\item The combination of looking to the past and to the future provides even more flexibility, giving the power to express exactly what is intended, for precisely satisfying some properties.
\end{itemize}
While the first two features where already present in the \textbf{AspectKB} framework (and in particular the first one is widely used in the aspect-orientation community), the third one is a very powerful add-on provided by the new \textbf{AspectKB+} framework.
\end{subsection}

\end{section}

\begin{section}{Conclusion}
\label{sec:Conclusion}
We have studied the problem of enforcing multilevel security in a distributed system as precisely as possible. An information-flow approach poses the problem of having to ``guess'' what processes in other locations may do, thereby losing some precision. Therefore, we have extended an existing framework to deal with a notion of localised state, which has given us the power to access information about the past performance of the system, thereby allowing us to capture the Bell-LaPadula policy with precision.

The resulting framework provides a way to combine policies that look to both the future and the past due to the four-valued Belnap Logic. This gives flexibility to the framework, by capturing precisely what is intended by the security policies. This also gives more power than the previous framework of \cite{AspectKB}.

\begin{paragraph}{Acknowledgements.} This work was partially funded by the Danish Strategic Research Council (project 2106-06-0028) ``Aspects of Security for Citizens'' and partially by the EU Integrated Project SENSORIA (contract 016004). We would like to thank Alan Mycroft for his comments on an early version of this paper. Finally, we really appreciated the comments from all the reviewers, they were very helpful.
\end{paragraph}

\end{section}

\bibliographystyle{eptcs}
\bibliography{references}

\end{document}